\documentstyle[pre,aps,preprint,psfig]{revtex}

\tightenlines

\begin{document}

\author{Felix von Oppen,$^1$ 
Bertrand I.\ Halperin,$^2$ Steven H.\ Simon,$^3$ and Ady Stern$^4$ }

\title{The half-filled Landau level -- composite fermions and dipoles}

\address{
$^1$Institut f\"ur Theoretische Physik, Universit\"at zu K\"oln, 
50937 K\"oln, Germany\\
$^2$ Physics Department, Harvard University, Cambridge, Massachusetts 02138,
USA \\
$^3$ Lucent Technologies Bell Labs, Murray Hill, NJ 07974, USA\\
$^4$ Department of Condensed Matter Physics, The Weizmann
Institute of Science, 76100 Rehovot, Israel}

\maketitle

\begin{abstract}
  The composite-fermion approach as formulated in the Fermion
  Chern-Simons theory has been very successful in describing the
  physics of the lowest Landau level near Landau level filling factor
  $\nu=1/2$. Recent work has emphasized the fact that the true low
  energy quasiparticles at this filling factors are electrically
  neutral and carry an electric dipole moment. In a previous work, we
  discussed at length two formulations in terms of dipolar
  quasiparticles.  Here we briefly review one approach -- termed
  electron-centered quasiparticles -- and show how it can be extended
  from $\nu=1/2$ to nearby filling factors $\nu$ where the low-energy
  quasiparticles carry both an electric dipole moment and an overall
  charge $e^*=(2\nu-1)e$.
\end{abstract}

\section{Introduction}

The ``composite fermion'' picture \cite{CFgeneral,sarma} has been very
useful for understanding many aspects of the fractional quantum Hall
effect (FQHE).  Jain \cite{Jain} showed that the most prominent FQHE
plateaus at Landau level filling factors $\nu=p/(2p+1)$, with $p$ an
integer, can be understood as the integer quantum Hall effect of
composite fermions and proposed approximate but extremely good trial
wavefunctions for these states. These wavefunctions can be thought of
as binding two vortices (zeros) of the wavefunction to each electron,
turning it into a ``composite fermion.'' Using a fermion Chern Simons
(FCS) theory, a field theoretic approach was developed
\cite{Lopez,Others} that is closely related to Jain's wavefunction
picture.  The FCS approach was used by Halperin, Lee, and Read
\cite{HLR} (HLR) to describe phenomena at, or near,
even-de\-no\-mi\-na\-tor filling fractions, such as $\nu = 1/2$, where
quantized Hall plateaus are not observed.  The most striking outcome
of this work was that the even-denominator fractions are {\it
  compressible and Fermi-liquid-like}.

The FCS approach is usually formulated in terms of bare quasiparticles
consisting of electrons to which two {\it fictitious} magnetic flux
quanta are attached by a singular gauge transformation. Thus, the bare
CS fermions carry charge $-e$. It was noted by Read \cite{NickRead}
that the actual low-energy quasiparticles at $\nu=1/2$ are in fact
electrically neutral due to screening by the magnetoplasmon modes and
that in addition, they carry an electric dipole moment perpendicular
to their canonical momentum. While several features of the FCS theory
had attractive interpretations in terms of dipolar quasiparticles, the
precise relationship to the FCS theory remained unclear at first.

This motivated a number of recently proposed theories of the
half-filled Landau level
\cite{Shankar,Pasquier,ReadNew,DHLee,Comment,prb,ShankarNew}, which
make the dipolar nature of the low-energy quasiparticles explicit.
Shankar and Murthy \cite{Shankar} also use their approach to derive
results for filling factors away from $\nu=1/2$. A review of all of
these approaches is beyond the scope of this article.  Instead, we
focus on some aspects of the work in Ref.\ \cite{prb}.

In the following, we first review the principal predictions of
the Fermion-Chern-Simons approach \cite{HLR}. In Sec.\ 3, we 
discuss the dipole picture \cite{NickRead}, stressing an apparent
discrepancy with the FCS approach and its resolution. We will then
present in more detail one particular way of obtaining dipolar
quasiparticles from the FCS theory \cite{prb} and show how it can be
extended to filling factors away from $\nu=1/2$.

\section{The fermionic Chern-Simons approach}

Attaching two fictitious flux quanta to the electrons by a unitary
transformation, one finds the Hamiltonian \cite{HLR}
\begin{equation}
\label{hamiltonian}
H=\int d{\bf r}{1\over 2m}\psi^\dagger(-i\nabla+e{\bf A} -e{\bf
  a})^2\psi+{1\over2}\int d{\bf r}\,d{\bf r^\prime} \psi^\dagger({\bf
  r})\psi^\dagger({\bf r^\prime}) v({\bf r}-{\bf r^\prime})\psi({\bf
  r^\prime})\psi({\bf r}),
\end{equation}
where $\psi$ denotes the field operator of the (bare) CS fermion,
${\bf A}$ the externally applied magnetic field and $v$ the Coulomb
interaction.  Due to the attached flux quanta, the CS fermions
interact not only via the usual Coulomb repulsion but also via a
Chern-Simons gauge field ${\bf a}$ which is subject to the
constraint
\begin{equation}
\label{csconstraint}
  {\bf \nabla}\times {\bf a}=2\phi_0\psi^\dagger\psi.
\end{equation} 
Here $\phi_0=h/e$ is the flux quantum.  In Eq.\ (\ref{hamiltonian}),
we chose the CS field in the Coulomb gauge $\nabla\cdot{\bf a}=0$.

The CS fermions are subject to an effective magnetic field $\Delta
B=\nabla\times({\bf A}-{\bf a})$ to which they couple like particles
of charge $-e$. At half filling, the fictitious magnetic field
associated with ${\bf a}$ cancels the externally applied magnetic
field on average ($\psi^\dagger\psi\to n$ with $n$ the electron
density).  As a result, the quasiparticles can travel in straight
lines over large distances, oblivious to the effects of the strong
applied magnetic field.  At filling fractions slightly away from
$\nu=1/2$, the cancellation is no longer perfect and the
quasiparticles should move in a circle, with a radius given by the
effective cyclotron radius $R^*_c = \hbar k_f/e|\Delta B|$ \cite{HLR}.
This prediction has been confirmed by several experiments
\cite{Experiments}. (Here $k_f$ is the Fermi-momentum, related to the
electron density by $k_f = (4 \pi n)^{1/2}$.)

Electronic response functions have been computed from the FCS theory
in random-phase approximation at various levels of sophistication
\cite{HLR,SimonHalperin}. In particular, the electronic (Coulomb
irreducible) density-density response function at $\nu=1/2$, for
$q,\omega$ small and $\omega\ll v^*_F q$, becomes
\begin{equation}
\label{rhorho}
  \Pi^e_{\rho\rho}({\bf q},\omega)={1\over{2\pi\over m^*}+{(2\phi_0)^2
    \over24\pi m^*}-i(2\phi_0)^2{2n\omega\over k_fq^3}}.
\end{equation}
This expression includes various central predictions of HLR \cite{HLR}:
\begin{itemize}

\item Very important in the present context is the prediction that the
  state at filling factor $\nu=1/2$ is {\it compressible}.  This means
  that the static limit of $\Pi^e_{\rho\rho}$ remains finite when
  $q\to 0$. Since $\Pi^e_{\rho\rho}$ describes the electronic density
  response to the (total) electric potential,
  $\Delta\rho^e=\Pi^e_{\rho\rho}\phi$, this implies that a static change
  in the electric potential produces a change in the electron density
  at arbitrarily long wavelengths.

\item The longitudinal conductivity $\sigma_{ll}({\bf q},\omega)
  =(i\omega/q^2)\Pi^e_{\rho\rho}$ is linear in the wavevector ${\bf q}$
  for $\omega\ll v^*_F q$. This has been confirmed in surface-acoustic
  wave experiments by Willett {\it et al.} \cite{Experiments}.

\item Charge relaxation is very slow at $\nu=1/2$ with $\omega \propto
  iq^2$. [This follows from the corresponding Coulomb reducible
  density response function which has an additional $v(q)$ in the
  denominator compared to (\ref{rhorho})]. The coupling to this slow
  mode, via the transverse CS field, leads to a logarithmic divergence
  in the effective mass in the single-particle Green functions. The
  effective mass $m^*$ entering response functions such as
  $\Pi^e_{\rho\rho}$ at $\nu=1/2$ is expected to be finite
  \cite{SternHalperin,DivergencesCancel}.

\end{itemize}
In addition, $\Pi^e_{\rho\rho}$ has a pole at the bare cyclotron
frequency $\omega_c$ for $\omega\gg v^*_Fq$, as mandated by Kohn's
theorem.

\section{Dipolar quasiparticles}

It is well known that the quasiparticle charge at the principal
fractional quantized Hall states at filling factors $\nu=p/(2p+1)$ is
equal to $e^*=e(2\nu-1)=-e/(2p+1)$. As we approach half filling
$\nu=1/2$ for $p\to\infty$, one finds that $e^*\to0$.  This allows for
a different interpretation of the effective cyclotron radius of the
bare composite fermions \cite{NickRead}: A quasiparticle with charge $e^*$
which sees the full magnetic field $B$ will have the same effective
cyclotron radius $R^*_c$ as a bare composite fermion of charge $-e$
that sees the effective field $\Delta B$.

Indeed, Read \cite{NickRead} noted that the true low-energy
quasiparticles in the fermion Chern-Simons theory, obtained upon
screening by the magnetoplasmon mode, are overall electrically
neutral: The introduction of a CS fermion into the system goes along
with a time-dependent flux which induces an electric field azimuthally
around the CF. Due to the finite Hall conductivity, this results in a
net flow of charge away from the CS fermion. In the adiabatic limit,
one finds that the overall charge flowing to infinity equals $-2\nu
e$.  Thus, the two flux lines can be viewed as a vortex of {\it
  positive} charge $2\nu e$ whose electrostatic attraction to
electrons leads to the formation of composite fermions
\cite{NickRead}.

Based on a trial wave function for $\nu=1/2$, Read \cite{NickRead}
observed that vortex and electron are separated from one another by a
distance proportional and perpendicular to the canonical momentum
${\bf k}$ of the low-energy quasiparticles \cite{SimonReview}. Thus,
the latter carry an electric dipole moment $ e\ell^2{\bf\hat
  z}\times{\bf k}$.  ($\ell$ is the magnetic length.)  Due to the
particular form of the dipole moment, a space dependent momentum
density ${\bf g}$ of the dipoles is associated with a charge density
by
\begin{equation}
\label{dipole-density}
   \rho^e=-{1\over B}\nabla\times{\bf g}.
\end{equation}
The Coulomb energy of the composite object grows with increasing
separation and hence with ${\bf k}$, giving a rationale for a Coulomb
generated effective mass.  Moreover, a dipole in a perpendicular
magnetic field performs ${\bf E}\times {\bf B}$ drift which provides
an alternate view of why excitations at $\nu=1/2$ propagate along
straight lines despite the large external magnetic field.

One may now be tempted to describe the low-energy physics of the
half-filled Landau level as a conventional Fermi liquid of dipolar
quasiparticles. However, this leads to a paradox, first noted by
Murthy and Shankar \cite{Shankar}: Dipoles couple only to gradients of
electric fields.  In particular, the coupling to electric potentials
becomes progressively weaker as the wavelength increases.
Hence, a conventional Fermi liquid of dipoles
would {\it not} be compressible, contrary to the predictions of HLR
\cite{HLR}. Technically, such an approach would predict that the
zero-frequency density density response function $\Pi_{\rho\rho}^e$ at
$\nu=1/2$ vanishes $\propto q^2$ in the limit $q\rightarrow 0$, in
disagreement with Eq.\ (\ref{rhorho}) above.

The resolution of this paradox \cite{Comment,prb} lies in a peculiar
symmetry, noted by Haldane \cite{Haldane}, of the system of dipolar
fermions at $\nu=1/2$: {\it The total energy of this particular system
  of dipoles remains unchanged if a constant ${\bf K}$ is added to the
  momentum of every particle.} Clearly, such a behavior is very
different from conventional Fermi liquids for which such a boost would
be expensive in terms of kinetic energy.  As a consequence of this
${\bf K}$ invariance, it costs very little energy to produce a long
wavelength fluctuation in the transverse momentum density $g_t$ of the
dipolar particles and the correlator $\langle g_t({\bf
  q},\omega=0)g_t(-{\bf q},\omega=0) \rangle$ diverges at small ${\bf
  q}$ as $1/q^2$.  In view of the relation (\ref{dipole-density}), one
then finds
\begin{equation}
   \Pi^e_{\rho\rho}({\bf q},\omega=0)\propto q^2\langle g_t({\bf q},\omega=0)
    g_t(-{\bf q},\omega=0)\rangle\propto {\rm const},
\end{equation}
implying that the electronic system is compressible. We will see in the
next section how ${\bf K}$ invariance is related to the underlying 
gauge symmetry of the system.

\section{Electron-centered quasiparticles near $\nu=1/2$}

In this section, we discuss in more detail one possible way of making
the dipolar quasiparticles explicit, starting from the FCS theory
\cite{prb}.  This approach, termed ``electron-centered
quasiparticles'' in Ref.\ \cite{prb}, is particularly simple and close
in spirit to the original approach of HLR. We use this opportunity to
show how it can be extended to filling factors $\nu$ away from $1/2$
where the quasiparticles are expected to carry both a dipole moment
and an overall electric charge $e^*=2\nu-1$.

The Chern-Simons action for $\nu=1/2$, in gauge-invariant form, is
\begin{eqnarray}
\label{basicaction}
S&=&\int dt\,d{\bf r}\left\{{1\over 8\pi}
  \epsilon_{ijk}a_i\partial_j a_k +\bar\psi
  i\partial_0\psi-a_0(\bar\psi\psi-n)\right.  \nonumber\\ 
&&\,\,\,\,\,\,\,\,\,\,\,\,\,\,\,\,\,\,\,\,\,\,\left.-{1\over
    2m}\bar\psi (-i\nabla+{\bf A}_{\rm eff}-{\bf
    a})^2\psi\right\}+S_{\rm Coulomb}.
\end{eqnarray}
Here, $\epsilon_{ijk}$ is the totally antisymmetric tensor and $\psi$
denotes the field of the Chern-Simons fermions. In this section, we
use units in which the electronic charge is -1 and $\hbar=1$.  We
already absorbed part of the external vector potential ${\bf A}$
corresponding to the applied magnetic field into the Chern-Simons
field, such that $\langle {\bf a}\rangle=0$ and ${\bf A_{\rm
    eff}}={\bf A}/(1-2\nu)$.  Using the constraint inherent in the
Chern-Simons theory, the Coulomb interaction can be written in terms
of the CS field,
\begin{equation}
  S_{\rm Coulomb}=-{1\over32\pi^2}\int dt\,d{\bf
    r}\,d{\bf r^\prime}[\nabla\times {\bf a}({\bf r})] v({\bf r}-{\bf
    r^\prime})[\nabla^\prime\times{\bf a}({\bf r^\prime})].
\end{equation}
HLR \cite{HLR} worked in the Coulomb gauge in which $a_0$ becomes a
Lagrange multiplier field enforcing the constraint
(\ref{csconstraint}). Recently, Murthy and Shankar \cite{Shankar}
observed that the CS field becomes dynamic in the temporal gauge
$a_0=0$ and describes the magnetoplasmon oscillators at $\nu=1/2$.
Working also in the temporal gauge, we derived a purely fermionic
low-energy, long wavelength action by integrating out the Chern-Simons
field \cite{prb}. The expression for the charge current and density in
terms of the fermionic fields entering this action allowed us to
identify the fermions as dipoles with the expected properties.

Here, we consider the CS theory in the more general gauges
\begin{equation}
\label{gauge}
   \lambda\nabla a_0=(1-\lambda)\partial_0{\bf  a_l},
\end{equation}
where $a_0$ denotes the scalar CS potential and ${\bf a_l}$ denotes the
longitudinal component of the CS vector potential ${\bf a}$. (The
transverse component of ${\bf a}$ will be denoted by ${\bf a_t}$.)  With
appropriate boundary conditions, we recover the Coulomb gauge
$\nabla\cdot{\bf a}=0$ for $\lambda=0$, while $\lambda=1$ corresponds
to the temporal gauge $a_0=0$.  

The reason for this more general gauge choice is the following. In the
limit of large magnetic field or, equivalently, vanishing band mass,
the problem at hand has a clear separation of scales. One the one
hand, there are inter-Landau-level processes at energies of the order
of the cyclotron energy $\hbar\omega_c$. On the other hand, the scale
of intra-Landau level processes is given by typical (Coulomb)
interaction energies. The dipolar excitations, being low-energy
quasiparticles, are expected to describe only the {\it intra-Landau
  level} physics. It turns out that this separation of scale is in
general {\it not} manifest in the FCS action. The gauge choice
(\ref{gauge}) is a device to make this separation of scales explicit
in the action and to derive an effective action for the low-energy
quasiparticles. The prediction for physical quantities should of
course be independent of the choice of gauge.

Using the gauge (\ref{gauge}), we can express the Chern-Simons term in
(\ref{basicaction}) in terms of the vector potential ${\bf a}$ alone,
\begin{equation}
  {1\over 8\pi}\epsilon_{ijk}a_i\partial_j a_k \to
  {1\over 4\pi\lambda}a_l \partial_0 a_t.
\end{equation}
This implies that $a_l$ and $a_t$ are canonically conjugate and the
corresponding operators satisfy the commutation relation
$[a_l,a_t]=-i4\pi\lambda$.  The presence of $\lambda$
indicates that the dynamics of the Chern-Simons field depends on the
gauge. In fact, ignoring the coupling to the fermions and the Coulomb
interaction for the moment, we find from (\ref{basicaction}) that the
Hamiltonian for the Chern-Simons field is $H=(n/2m)(a_l^2+a_t^2)$.
Together with the commutation relations, we deduce that the
Chern-Simons field describes oscillators of frequency
\begin{equation}
  \omega_\lambda={4\pi\lambda n\over m}.
\end{equation}
We can now fix $\lambda$ by demanding that the oscillators describe
the magnetoplasmon mode at the cyclotron frequency $\omega_c=B/m$.
This implies
\begin{equation}
\label{gaugechoice}
   \lambda={1\over2\nu}.
\end{equation} 
For $\nu=1/2$, we recover the temporal gauge $\lambda=1$ of Refs.\ 
\cite{Shankar,prb}.  We will now adopt the choice of gauge
(\ref{gaugechoice}) and derive an effective action for the low-energy
quasiparticles.

In the following, it will be convenient to define the 
(gauge-dependent) momentum current
\begin{equation}
  {\bf g}={1\over 2}\left\{\bar\psi(-i\nabla+{\bf A}_{\rm eff})\psi
    +[(i\nabla+{\bf A}_{\rm eff})\bar\psi]\psi\right\}.
\end{equation}
Furthermore defining
\begin{equation}
  {\bf \tilde g}={\bf g}+{m\omega\over q}{1-\lambda\over \lambda}\rho
  {\bf \hat q} 
\end{equation}
with $\rho({\bf r},t)=\bar\psi({\bf r},t)\psi({\bf r},t)-n$ and
\begin{equation}
  U_\lambda^{-1}=\left(\begin{array}{cc} 0 & -i\omega/
      [4\pi\lambda] \\ i\omega/[4\pi\lambda] &
      -q^2v(q)/(4\pi)^2 \end{array}\right),
\end{equation}
we have for the action
\begin{eqnarray}
  S=-{1\over 2}{\bf a}(n/m-U_\lambda^{-1}){\bf a}+\bar\psi
  i\partial_0\psi -{1\over 2m}\bar\psi(-i\nabla+{\bf A}_{\rm
    eff})^2\psi+{1\over m} \tilde{\bf g}\cdot{\bf a}.
\end{eqnarray}
Here, we neglect the fluctuations of the density in the diamagnetic
term.  Integrating out the Chern-Simons field ${\bf a}$, we obtain
\begin{eqnarray}
  S=\bar\psi i\partial_0\psi-{1\over 2m}\bar\psi(-i\nabla+{\bf A}_{\rm
    eff})^2\psi
     +{1\over 2m^2}\tilde{\bf g}\,{1\over n/m-U_\lambda^{-1}}\,
    \tilde{\bf g}.
\end{eqnarray}

A low energy, long wavelength action can now be derived by noting that
the entries of the matrix $U_\lambda^{-1}$ are proportional either to
frequency $\omega$ or momentum ${\bf q}$. While this is true for all
gauges $\lambda$, the expansion is in terms of $\omega/\omega_c$ (at
$q=0$) as expected on physical grounds for the choice
(\ref{gaugechoice}) only. Following Ref.\ \cite{prb}, we keep all
terms which remain finite in the limit $m\to 0$. (Remember that the
limit of vanishing band mass is equivalent to the limit of high
magnetic field.) This gives the effective action
\begin{equation}
\label{low-e-action}
S=\bar\psi i\partial_0\psi-{1\over 2m}\bar\psi(-i\nabla+{\bf A}_{\rm
  eff})^2\psi +{1\over 2mn}\tilde{\bf g}\cdot\tilde{\bf g} +{1\over
  2n^2}{\bf g}U_\lambda^{-1}{\bf g}.
\end{equation}
For a physical interpretation of this result, it is useful to
introduce a source field ${\bf\cal A}$ by $\tilde{\bf g}\to\tilde{\bf
  g}+n{\bf\cal A}$ and ${\bf A}_{\rm eff}\to{\bf A}_{\rm eff}+{\bf\cal
  A}$ so that the expression for the physical charge current ${\bf
  j^e}$ follows from ${\bf j^e}=\partial S/\partial {\bf\cal A}$. One
readily finds from Eq.\ (\ref{low-e-action})
\begin{equation}
\label{current}
{\bf j^e}={1\over n}U_\lambda^{-1}{\bf g}+{\omega\over q}
(2\nu-1)\rho{\bf \hat q}.
\end{equation}
Using the electronic continuity equation, we obtain for the charge
density
\begin{equation}
\label{density}
\rho^e=(2\nu-1)\rho-{1\over B}iqg_t.
\end{equation}
Here the second term was already found at $\nu=1/2$ in \cite{prb}. It
shows that in this theory, a fermion of momentum ${\bf k}$ carries a
dipole moment $-(1/B){\bf \hat z}\times {\bf k}$. The first term is
nonzero only away from $\nu=1/2$ and shows that the fermions also
carry charge $e^*=2\nu-1$, as expected for the true
low-energy quasiparticles.

The action (\ref{low-e-action}) combined with the expressions
(\ref{current}) and (\ref{density}) for the charge current and charge
density are the desired result. Several comments are in order.

1) We are now able to compute the electronic density-density
correlator in two different ways. On the one hand, we can use that
$\rho^e =-(\bar\psi\psi-n)$. In addition, we have the expression
(\ref{density}). This leads to the consistency condition
\begin{equation}
  -\rho=(2\nu-1)\rho-{1\over B}iqg_t,
\end{equation}
which simplifies to
\begin{equation}
\label{consistency}
\rho={1\over4\pi n}iqg_t.
\end{equation}
In fact, precisely the same consistency condition was already found at
$\nu=1/2$ \cite{prb}. In this formulation, it is obvious that this
consistency condition is closely related to the constraint in the
Hamiltonian approach of Ref.\ \cite{Shankar}.  

One may ask whether the action (\ref{low-e-action}) satisfies this
consistency condition. It turns out that this is closely related to
the analog of the continuity equation for the action
(\ref{low-e-action}). As usual, the invariance of the action under the
(global) infinitesimal transformation $\psi\to\psi+i\alpha\psi$ and
$\bar\psi\to\bar\psi-i\alpha\bar\psi$ implies a conservation law at
the semiclassical level.  For standard fermionic actions, this is just
the continuity equation. A simple calculation shows that the analogous
conservation law for the action (\ref{low-e-action}) just implies that
the consistency condition is a constant of motion,
\begin{equation}
  i\omega\left\{\rho-{1\over4\pi n}iqg_t\right\}=0.
\end{equation}
This shows that the action satisfies the consistency condition.  We
note that the last term in the action (\ref{low-e-action}) is crucial
for this to work.  Using this constant of motion also gives the
simpler expression
\begin{equation}
\label{density-simple}
\rho^e=-{1\over4\pi n}iqg_t.
\end{equation}
for the charge density. 

2) It is interesting to take a closer look at the last term in the
action (\ref{low-e-action}). Multiplying this term out, we find two
contributions.  One contribution involves the Coulomb interaction
$v(q)$. In fact, using Eq.\ (\ref{density-simple}), one observes that
it is simply a rewriting of the Coulomb term in the action.

In addition, there are terms due to the off-diagonal entries in
$U_\lambda^{-1}$, of the structure ${\bf g}\times\partial_0{\bf g}$.
Besides being important for satisfying the consistency condition
(\ref{consistency}) as mentioned above, these terms have another
interesting function. Due to the time derivatives, these imply that in
a corresponding Hamiltonian description, the fields $\psi$ and
$\bar\psi$ are no longer canonically conjugate and their
anticommutator differs from the canonical one. Following standard
procedures for the quantization of constrained systems \cite{Pauli},
one can derive the modified anticommutator (Dirac bracket) of $\psi$
and $\bar\psi$. Interestingly, it turns out that the modification is
such that the commutator of the electronic density
$\rho^e=-(\bar\psi\psi-n)$ now reproduces the lowest-Landau level
commutator at long wavelengths \cite{unpublished}.

3) The low-energy action (\ref{low-e-action}) contains current-current
interaction terms. {\it These interactions are crucial for the action
  to satisfy ${\bf K}$ invariance.} If there were only the
kinetic-energy term in the action, a boost of all fermions by the
momentum ${\bf K}$ would change the action $S_{\rm kin}\to S_{\rm kin}
-({\bf K}/m)\cdot g({\bf q}=0,\omega=0)-nK^2/2m$. The role of the
attractive current-current interaction ${\bf g}\cdot{\bf g}/2mn$ is to
cancel precisely this cost in kinetic energy. Noting that boosting all
fermions by ${\bf K}$ amounts to shifting ${\bf g}({\bf
  q}=0,\omega=0)\to{\bf g}({\bf q}=0,\omega=0)+n{\bf K}$, we find that
the action (\ref{low-e-action}) is indeed ${\bf K}$ invariant. 

The origin of ${\bf K}$ invariance in this formulation can be traced
back to gauge invariance. Even once we specify in the FCS action
(\ref{basicaction}), e.g., to the temporal gauge $a_0=0$, we can still
make the limited gauge transformations
\begin{eqnarray}
  a_l({\bf q},\omega=0)&\to& a_l({\bf q},\omega=0) + f({\bf q})
  \nonumber\\
  a_t({\bf q}=0,\omega=0)&\to& a_t({\bf q}=0,\omega=0) + {\rm const}.
\end{eqnarray}
These transformations must leave the physical current unchanged,
provided that we simultaneously change the canonical momentum 
density of the fermions by
\begin{eqnarray}
\label{kinvariance}
  g_l({\bf q},\omega=0)&\to& g_l({\bf q},\omega=0) + nf({\bf q})
  \nonumber\\
  g_t({\bf q}=0,\omega=0)&\to& g_t({\bf q}=0,\omega=0) + n\,{\rm const}.
\end{eqnarray}
For $q=0$ this transformation shifts the momentum of each fermion by a
constant. Above, we eventually integrate out the CS field ${\bf a}$
after which the action must be invariant under the transformation
(\ref{kinvariance}) alone. This is precisely what we call ${\bf K}$
invariance. The relation to gauge invariance emphasizes the fundamental
role of ${\bf K}$ invariance for the present system.

4) At $\nu=1/2$ one can now readily compute the electronic response
functions in random-phase approximation. In this way, one recovers 
Eq.\ (\ref{rhorho}) precisely. 

5) As for the original CS fermions, the effective mass of the
single-particle Green function of the dipolar quasiparticles diverges
logarithmically at $\nu=1/2$ (for Coulomb interaction) \cite{prb}.  In
the present formulation, this is due to the singular behavior of the
correlator of $g_t$. This can be understood by noting that the
divergence in the quasiparticle mass arose in HLR \cite{HLR} from
coupling the CS fermions to the transverse CS field which remains
unaffected by gauge transformations.

\section{Conclusions}

The present paper is a brief review of some results obtained in a
recent paper \cite{prb}. In addition, we extended the approach in
terms of ``electron centered quasiparticles'' to filling factors away
from $\nu=1/2$, where the low-energy quasiparticles carry not only a
dipole moment but also an electric charge. This charge agrees with 
the quasiparticle charges expected at the principal quantum Hall
states $\nu=p/(2p+1)$.  Due to the brevity of the present paper, many
important points remained unmentioned. We use these conclusions to
list a few of them.

\begin{itemize}
 
\item The quasiparticles of the previous section are not the only
  possible ones. Indeed, we find \cite{prb} that there are several
  different ways of defining quasiparticles which differ in the
  long-wavelength limit but which nevertheless predict the same
  electronic response functions.  Physically, this freedom is
  associated with different definitions of the position of the
  quasiparticle.  Due to the relation $\rho^e=-(\bar\psi\psi-n)$, the
  coordinates of the dipoles as defined in the previous section
  coincide with the electronic positions. On the other hand, the
  Hamiltonian approach of Murthy and Shankar \cite{Shankar} leads to
  dipoles whose position is defined to be half way between the
  electrons and the vortices \cite{prb}.

\item Strictly speaking, the calculations of the previous section are
  valid only in a model in which the charge and flux of the CS Fermion
  are spread over a finite radius $Q^{-1}$.  In fact, we argue that
  the random phase approximation becomes exact in the limit of small
  $Q$ \cite{prb}. This model also enabled us \cite{prb} to carry out
  explicitly the Murthy-Shankar unitary transformation from CS
  fermions to dipole fermions. The Hamiltonian we obtain for the
  dipole fermions is not that of free dipolar particles, but rather a
  more complicated one, and also reproduces the response function
  originally predicted by HLR.

\item We pass from the small $Q$ to the physical case of unlimited $Q$
  by showing that the assumption of ${\bf K}$ invariance together with
  the dipole relation (\ref{dipole-density}) and conventional
  Fermi-liquid assumptions is bound to lead to an electronic
  density-density response function of the form (\ref{rhorho}).  In
  particular, the static limit of this response function predicts the
  $\nu=1/2$ state to be compressible.

\end{itemize}

{\bf Acknowledgments:} This research was supported in part by the
National Science Foundation under Grants No. PHY94-07194 and
DMR-94-16910 (BIH), by the US-Israel Binational Science Foundation
(95-250) (BIH and AS), by the Minerva foundation (FvO and AS), by the
Israel Academy of Science, DIP, and the V.\ Ehrlich career development
chair (AS), and by SFB 341 K\"oln-Aachen-J\"ulich (FvO).


\begin{thebibliography}{99}

\bibitem{CFgeneral} For a general review of composite fermion physics,
  see {\it Composite Fermions}, ed. O. Heinonen, (North Holland,
  1998).

\bibitem{sarma} See also {\it Perspectives in Quantum Hall Effects},
  edited by S. Das Sarma and A. Pinczuk, (John Wiley and Sons, 1997)
 
\bibitem{Jain} J. K. Jain, Phys. Rev. Lett. {\bf 63}, 199 (1989); for
  a more recent discussion of Jain's approach see the chapter by J. K.
  Jain and R. Kamilla in Ref. \cite{CFgeneral} as well as J. K. Jain,
  Adv. Phys. {\bf 41}, 105 (1992).
  
\bibitem{Lopez} A. Lopez and E. Fradkin, Phys. Rev. B {\bf 44} 5246
  (1991); {\it ibid.} {\bf 47}, 7080 (1993); See also the chapter by
  A. Lopez and E. Fradkin in Ref. \cite{CFgeneral}.

\bibitem{Others} G. Moore and N. Read, Nucl. Phys. B {\bf 360}, 362
  (1991); M. Greiter and F. Wilczek, Mod Phys. Lett. {\bf B4}, 1063
  (1990); Nucl. Phys. B {\bf 370}, 577 (1992); M. Greiter, X.-G. Wen
  and F. Wilczek, Phys.  Rev. Lett. {\bf 66}, 3205; Nucl. Phys. B {\bf
    374}, 567 (1992).

\bibitem{HLR} B.~I.~Halperin, P.~A.~Lee, and N.~Read, Phys.  Rev. B
  {\bf 47}, 7312 (1993).

\bibitem{NickRead} N. Read, Semi. Cond. Sci. Tech. {\bf 9}, 1859
  (1994); Surf. Sci., {\bf 361/362}, 7, (1996).


\bibitem{Shankar} R. Shankar and G. Murthy, Phys. Rev. Lett.  {\bf
    79}, 4437 (1997).  See also the chapter by G. Murthy and R.
  Shankar in Ref. \cite{CFgeneral}.  
    
\bibitem{Pasquier} V. Pasquier and F. D. M. Haldane, Nucl. Phys. B
    {\bf 516} 719 (1998).

\bibitem{ReadNew} N. Read, cond-mat/9804294. 

\bibitem{DHLee} D.-H. Lee, Phys. Rev. Lett. {\bf 80} 4547 (1998).
  
\bibitem{Comment} B. I. Halperin and A. Stern, Phys. Rev. Lett. {\bf
    80}, 5457 (1998).

\bibitem{prb} A.\ Stern, B.I.\ Halperin, F.\ von Oppen, S.\ Simon,
  Phys.\ Rev.\ B, in press and cond-mat/9812135.

\bibitem{ShankarNew} R.\ Shankar, cond-mat/9903064. 

\bibitem{Experiments} A review of experimental results on composite
  fermion physics is given by R. L. Willett in Ref.  \cite{CFgeneral}.
  See also, R. L. Willett, Adv.  Phys. {\bf 46}, 447 (1997); and H. L.
  Stormer and D. C. Tsui, in Ref.  \cite{sarma}, Chap. 10; See also M.
  P. Lilly {\it et al.}, Phys.  Rev. Lett. {\bf 80}, 1714 (1998) for
  coulomb drag experiments; See J. P. Eisenstein, L. N. Pfieffer, and
  K. W. West, Phys. Rev. B {\bf 50}, 1760 (1994) for compressibility
  measurements; See J. S. Moon {\it et al.}, Phys. Rev. Lett. {\bf
    79}, 4457 (1997) for magnetocapacitance measurements.

\bibitem{SimonHalperin} S.~H.~Simon and B.~I.~Halperin, Phys. Rev. B
  {\bf 48}, 17386 (1993); See also Ref. \cite{SimonReview} and S. H.
  Simon, J. Phys. Cond. Matt. {\bf 8}, 10127 (1996).


\bibitem{SternHalperin} A.  Stern and B. I.  Halperin,
  Phys. Rev. B {\bf 52}, 5890 (1995); Surf. Sci. {\bf 361}, 42 (1996).

\bibitem{DivergencesCancel} Y. B. Kim, A. Furusaki, X.-G. Wen, and P.
  A. Lee., Phys.  Rev. B {\bf 50}, 17917 (1994); Y.  B. Kim, P. A.
  Lee, X.-G. Wen and P. C. E. Stamp, {\it ibid} {\bf 51} 10779 (1995).
  
\bibitem{SimonReview} A recent discussion of the relation between the
  wavefunction approaches and the dipole picture is given by
  S. H. Simon in Ref. \cite{CFgeneral}. 

\bibitem{Haldane} F.D.M.\ Haldane, unpublished.

\bibitem{Pauli} S.J.\ Brodsky, H.C.\ Pauli, and S.\ Pinsky, Phys.\ 
  Rep.\ {\bf 301}, 299 (1998), appendix E.

\bibitem{unpublished} A. Stern, F. von Oppen, B. I. Halperin, and S.
  H. Simon, unpublished.

\end{thebibliography}
\end{document}